\newcommand{\CLASSINPUTbottomtextmargin}{1.97cm}
\documentclass[10pt, conference, compsocconf]{IEEEtran}
\usepackage{graphicx}
\usepackage{caption}
\usepackage[font=footnotesize]{caption}
\usepackage{subfig}
\usepackage{color}
\usepackage{flushend}

\usepackage{algorithm}
\usepackage{algpseudocode}

\algtext*{EndWhile}% Remove "end while" text
\algtext*{EndIf}% Remove "end if" text
\algtext*{EndFor}% Remove "end if" text

\renewcommand{\baselinestretch}{0.96}

\usepackage{xcolor}

\ifCLASSINFOpdf
\else
\fi
\begin{document}
\title{LUT-Lock: A Novel LUT-based Logic Obfuscation for FPGA-Bitstream and ASIC-Hardware Protection \vspace{-15pt}}


\author{\IEEEauthorblockN{Hadi Mardani Kamali, Kimia Zamiri Azar, Kris Gaj, Houman Homayoun  and Avesta Sasan}
\IEEEauthorblockA{Electrical and Computer Engineering Department \\
George Mason University,
Fairfax, VA 22030 \\ Email: \{hmardani, kzamiria, kgaj, hhomayou, asasan\}@gmu.edu \vspace{-8pt}}
}


%Telephone: (800) 555--1212, Fax: (888) 555--1212}


\maketitle

\begin{abstract}

In this work, we propose LUT-Lock, a novel Look-Up-Table-based netlist obfuscation algorithm, for protecting the intellectual property that is mapped to an FPGA bitstream or an ASIC netlist.  We, first, illustrate the effectiveness of several key features that make the LUT-based obfuscation more resilient against SAT attacks and then we embed the proposed key features into our proposed LUT-Lock algorithm. We illustrates that LUT-Lock maximizes the resiliency of the LUT-based obfuscation against SAT attacks by forcing a near exponential increase in the execution time of a SAT solver with respect to the number of obfuscated gates. Hence, by adopting LUT-Lock algorithm, SAT attack execution time could be made unreasonably long by increasing the number of utilized LUTs.

\end{abstract}


% \fi
\IEEEpeerreviewmaketitle

\vspace{-6pt}
\section{Introduction} \label{intro}

Hardware security has become a major concern for both FPGA and ASIC solutions. For the ASIC solution, the problem of hardware security resides in using untrusted parties in the manufacturing supply chain for economically driven reasons. Due to the high cost of building, operating, managing, and maintaining state-of-the-art silicon manufacturing facilities, many major U.S. high-tech companies have been always fabless or went fabless in recent years \cite{6926108}, which has led them to adopt to offshore fabrication. However, many offshore fabrication facilities are considered to be untrusted, and fabrication in untrusted fabs has introduced multiple forms of security threats into the supply chain including threats of overproduction, Trojan insertion, Reverse Engineering , IP theft, and counterfeiting \cite{6926108}.  

On the other hand, FPGAs are inherently more secure for their post-silicon reconfigurability. However, the FPGA hardware security relies on protected and non-intruded mapping of the intended bitstream into FPGAs. In certain cases, it is difficult to protect the bitstream both during the initial configuration in untrusted third-party systems as well as during remote and in-field reconfiguration \cite{7857187}. A successful attack may result in an unauthorized transfer of a bitstream to a third-party, reverse-engineering of the embedded netlist, injection of a hardware Trojan, and cloning or theft of embedded IPs \cite{7857187}\cite{6461919}. Although high-end FPGAs are typically equipped with bitstream encryption, there are many cases where encryption alone is not enough \cite{6461919}: (1) Not all FPGA families are equipped with implementations of cryptographic algorithms \cite{7736459, mardani2016aes}, especially for small and low-energy FPGAs. (2) When the power and delay overhead of bitstream encryption process is not tolerable, a developer may choose not using encryption. (3) Many FPGA-based products, to support new services or to enhance the existing ones, require frequent updates which mostly accomplished remotely. Despite the first time safely programming, for in-field updates and a remote upgrade, the encrypted bitstream and the keys are vulnerable to leakage \cite{7857187}. (4) After deployment, FPGAs are susceptible to physical attacks. The long-term in-field usage makes it possible for an attacker to extract the encryption keys via various side channel attack mechanisms \cite{6285774}. So, it is essential to implement additional defensive measures to prevent the usability of a leaked bitstream. Such threats validates the need for implementing a strong obfuscation to hide the bitstream.

In this paper we propose LUT-Lock, which obfuscates a netlist while embedding several key features that make the obfuscation a hard problem for state of the art attacks with particular attention to Satisfiability (SAT) Attacks. To develop this defense mechanism, we have identified several key features that increase the difficulty of obfuscation for SAT attacks. We illustrate how by utilizing each feature during the obfuscation, the SAT problem becomes harder. We propose LUT-Lock algorithm which combines all features, providing the best defense against SAT attacks.  

The rest of the paper is organized as follows: Section \ref{background} provides background on the logic obfuscation, and the use of SAT solvers for deobfuscation. Section \ref{LUTobfuscation} justifies the use of LUTs for obfuscating ASIC and FPGA solutions. Section \ref{strategy} explains various LUT-based obfuscation sub-algorithms proposed in LUT-Lock and justifies their effectiveness. Section \ref{setup} describes our experimental setup. Section \ref{results} presents our experimental results and discusses our findings. And finally, section \ref{conclusion} concludes the paper. 

\vspace{-2pt}
\section{Background} \label{background}
\vspace{-3pt}

\vspace{-3pt}
\subsection{Obfuscation}

Logic obfuscation is the process of hiding the functionality of a synthesized IP by building ambiguity by means of control and programmability into its netlist. Gate camouflaging and logic locking are two of the widely explored obfuscation schemes in ASICs  \cite{6881480}\cite{7128395}\cite{7479225}. The claim raised by such obfuscation scheme was that to break the obfuscation, the adversaries need to try a large number of inputs and key combinations to extract the correct key, whose time increases exponentially as the number of keys and inputs increases. Note that in ASIC solutions, the availability of scan chains (for DFT), allows an adversary to access combinational logic in each stage of a sequential circuit.

The strength of logic obfuscation was seriously challenged by attacks formulated using satisfiability solvers (SAT Attacks), which were able to break the prior methods of logic obfuscation within minutes \cite{7140252}\cite{el2015integrated}. The strength of this attack directed the attention of HW security researchers to architect harder obfuscation schemes that are more resilient to SAT attacks. SARLock \cite{7495588}, Anti-SAT \cite{xie2016mitigating}, And-Tree-Insertion (ATI) \cite{li2017provably}, CamoPerturb \cite{yasin2016camoperturb}, SFLL-HD$^{0}$ \cite{yasin2017provably}, and SRCLock \cite{roshanisefat2018srclock} are some of the obfuscation approaches that were proposed for this purpose. However further research proved that some of these obfuscation techniques are prone to other types of attacks such as simple removal attack after identification of these blocks using Signal Probability Skew (SPS) attacks \cite{7858346}, and approximate-SAT attacks.

LUT-based obfuscation has been previously visited by few researchers. The work in \cite{baumgarten2010preventing} suggest using LUTs for obfuscation and provides several replacement strategies to secure a netlist. However, the proposed mapping algorithms are not resilient against SAT attacks, and are only evaluated in terms of power, performance and area (PPA) overhead, while the claim on the security of these schemes is made solely base on inability to readout the content of LUTs after reverse engineering. The work in \cite{7544331} proposed a STT-LUT-based obfuscation with three different LUT placement algorithms. This work further focuses on PPA impact of their solution and illustrates that utilizing STT-based LUTs could reduce the PPA impact. However, the proposed solution does not consider its resiliency against SAT attack.

\subsection{SAT Attack}

Every obfuscated gate in a netlist could be represented by a \emph{Key Programmable Gate} (KPG). A KPG, based on its key input, could be configured to take any of $n$ different possible functionalities. In XOR and MUX based obfuscation, the XOR and MUX are already a key programmable cell, where the key input to an XOR gate or select input of a MUX are considered as key inputs. Other obfuscated gates could be easily transformed into a key programmable cell. For example, a camouflaged gate could be represented by a MUX, with each possible output column of a truth table taken as an input to the MUX and the select inputs of a MUX used as key inputs. In SAT attack, the obfuscated netlist is first updated by converting all obfuscated cell to KPG cells. Let us refer to this circuit by a \emph{Key Programmable Circuit} (KPC). A SAT attack on an obfuscated circuit is an iterative process of finding an input and two key values $K_1$ and $K_2$ for which a KPC produces two different results, where one of them is the expected output, that could be verified by comparing it to the output of a functional circuit (\emph{eval}). Such input is denoted as a Discriminating Input (DIP). The SAT solver then formulates an additional constraint that in the future iterations such that in addition to producing a different output for a new input, the two keys ($K_1$ and $K_2$) should also produce the same output for all previously found DIPs. This constraint makes sure that solver reduces the set of possible keys for a circuit in each iteration. The SAT solver exits when it can no longer find a new DIP. At this point, any key that produces the correct output for all previously found DIPs is the correct key \cite{7140252}\cite{el2015integrated}.

\section{LUT-based Obfuscation for ASIC and FPGA} \label{LUTobfuscation}

\subsection{LUT-based obfuscation in FPGA}

\begin{figure}[t]%
    \centering
    \vspace{-10pt}
    \subfloat[]{{\includegraphics[width=0.44\columnwidth]{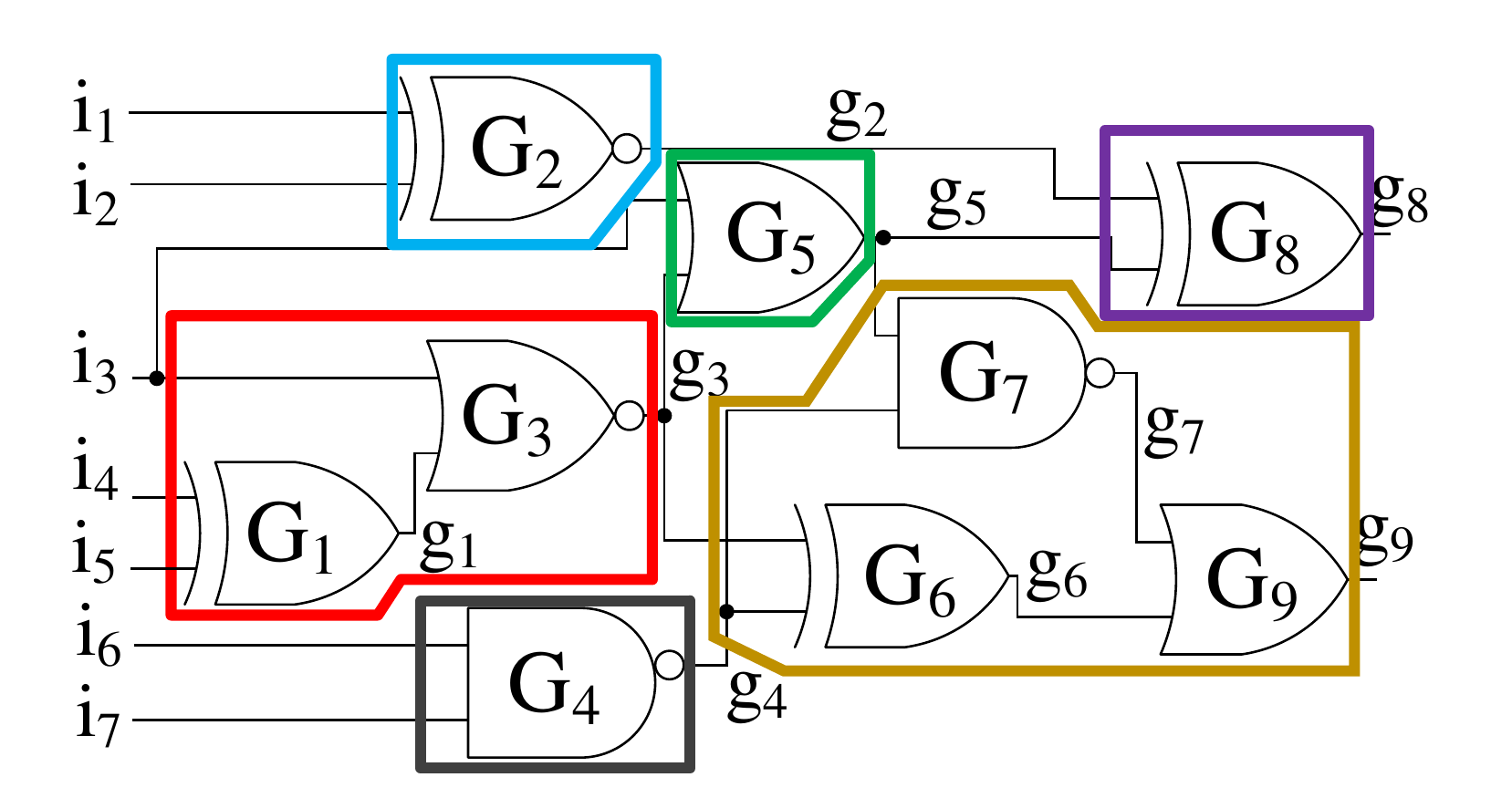} }}%
    \subfloat[]{{\includegraphics[width=0.44\columnwidth]{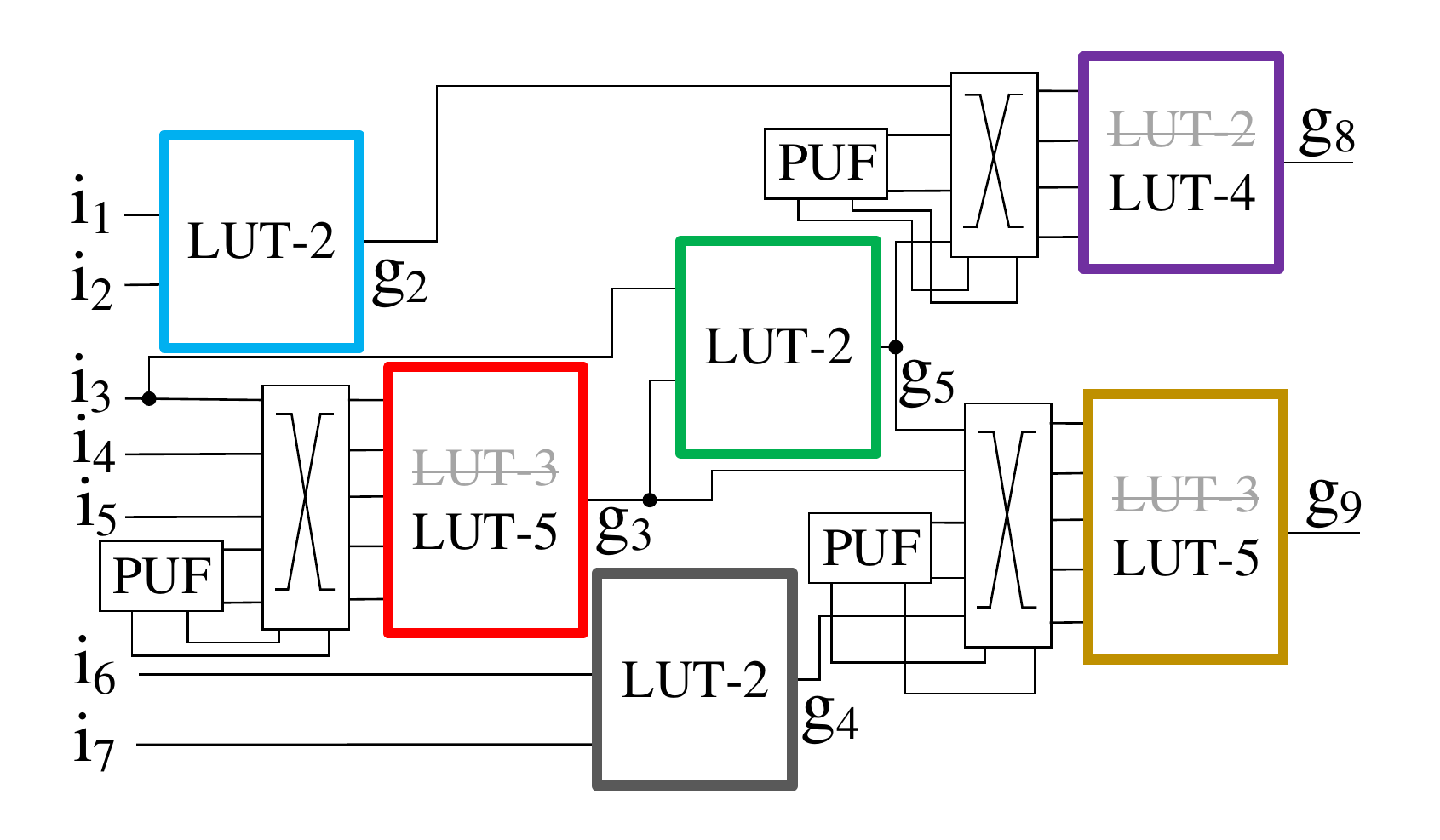} }}%
    \vspace{-4pt}
    \caption{(a) sample circuit (b) using configurable switches + PUFs (NLFSRs) for employing larger LUTs after synthesis.}%

    \label{LUTSynth}%
\end{figure}

In an FPGA solutions the hardware resources are are fixed and is designed independent of a given netlist. Hence by nature, state of the art FPGAs provide a large pool of resources to be applicable to a wide range of applications, resulting in a large number of non-utilized LUTs after mapping a netlist to the FPGA. For instance, the study in \cite{7857187} depicts the utilization of Altera Cyclone V after mapping a diverse set of benchmarks of various scale and complexity to this FPGA, and reported that FPGA utilization is typically low. This phenomenon was coined as FPGA-Dark-Silicon \cite{7857187}. These unmapped and unutilized LUTs are freely available and could be used for obfuscating a to-be-mapped netlist. Hence, LUT-based obfuscation in FPGAs could be considered as utilizing unused LUTs, or using larger than needed  LUTs, where the connectivity and impact of additional logic is controlled using keys. The process of using LUTs in FPGA for the purpose of logic obfuscation is illustrated in in Fig. \ref{LUTSynth}(b), where some of 2-input (or 3-input) logic gates could be mapped to a LUT of larger size (e.g. size 4 or 5). Then, the additional inputs can be taken from the output of an internally implemented Non-Linear Feedback Shift Register (NLFSR) or a Physical Unclonable Function (PUF) \cite{4261134}. In addition, by changing the ordering of inputs based on the key inputs (generated by PUF), the obfuscated circuit possibilities increases. Lets assume a PUF is used. In this case, each FPGA has a unique PUF response. By knowing the PUF response ahead of time, the bitstream will load the LUTs with proper values and will transmit the directives for connecting the known PUF outputs to the proper LUT inputs and switch box select lines. However, the PUF values will not be transmitted in the bitstream. This missing key values serve as the obfuscation key in LUT based obfuscation. Also note that the bitstream in this case is unique for each FPGA, as each FPGA has a unique PUF response. In this case, even if the bitstream is leaked, the PUF response remains unknown, making the problem similar to ASIC flow, where after reverse engineering the obfuscated netlist is available, but the keys are unknown. 

\subsection{LUT-based obfuscation in ASIC}

In ASICs, utilizing LUTs for obfuscation can lead to the considerable area and delay overhead. In the CMOS implementation of LUTs, the area overhead of the memory elements in a LUT exponentially increases as a function of its input size. Hence, the imposed area overhead limits the number of LUTs that could replace regular gates in a netlist. In addition, the performance/delay requirements constrain the placement of LUTs in timing critical and near timing critical paths. However, with the introduction of STT and MTJ based LUTs \cite{8074566}\cite{7544331} and the promise of integration of STT and MTJ/pMTJ-based LUTs into the same CMOS process, the area overhead of LUTs is expected to sharply reduce. Integration of CMOS and MTJ/STT devices makes it possible for a larger number of LUTs to be implemented given a fixed area overhead. Using LUTs for obfuscation in ASICs is straightforward: selected cells are removed and replaced by LUTs. The functionality of cell remains hidden to the manufacturer. LUTs are then programmed after fabrication in a trusted testing facility. 

\section{Proposed LUT-Lock Obfuscation Algorithm} \label{strategy}

Our proposed LUT-Lock algorithm combines several key features, each enhancing its ability to resist against SAT attacks. In this section, we first explain each key feature, and then propose the LUT-Lock algorithm that combines all features into a comprehensive solution. In the result section of this paper, we illustrate how by adding each key feature, the resiliency of obfuscated netlist against SAT attack increases, proving that the resiliency gained from adding this features are orthogonal to one another.

\subsection{FIC: Focusing on the Fan-In Cone of minimum number of primary output} 

The first criteria for selection of candidate gates is derived from the observation that higher output corruption reduces resiliency of obfuscation solution against SAT attacks \cite{7495588}\cite{xie2016mitigating}. Hence, by mapping the LUTs such that it affects the minimum number of primary outputs (POs), the degree of output corruption reduces, increasing the strength of obfuscation against SAT attacks. To achieve this, we limit the LUT insertion to the fan-in cone of smallest possible set of primary outputs (best case being single output), and we refer this algorithm as FIC. Note that FIC LUT-replacement still corrupts other outputs, as the intersection of fan-in cones of different outputs is not empty. In addition, the number of gates in the intersection of fan-in cones increases as we move from outputs toward inputs. Hence the obfuscation should be designed to replace the closest cells to the selected output first. This could be achieved by means of a Breadth First Search (BFS).  In order to avoid timing violation due to replacing a gate with LUT, we estimates the delay of all timing paths through a gate selected for replacement. If the estimated delay is more than predefined threshold (e.g. 10\% delay overhead), the allowance of replacement for this candidate will be revoked, and next candidate will be checked for replacement. After replacing all gates in the current Fan-In Cone, a new primary output will be selected.

In FIC algorithm, the output pin(s) selected for obfuscation should meet two conditions: (1) Total Positive Slack (TPS) of all timing paths leading to that primary output(s) should be large. This is because replacing a gate with LUT incurs additional delay in every timing path that passes through that gate. Hence, we need available timing slack for replacement of faster logic gates with slower LUTs. (2) it must have a large fan-in cone size, giving us more candidate gates for replacement. Fig. \ref{strategies}(a) illustrates the FIC replacement strategy. Between the two outputs, i.e. $g_8$ and $g_9$, $g_9$ is not selected, as it contains the largest number of timing critical paths. When using BFS for gate selection, FIC selects gates \{$G_8$ and $G_5$\} or \{$G_8$, $G_5$, $G_2$, and $G_4$\} when its asked to replace 2 or 4 gates respectively. For large circuits, we define two coefficients ($\alpha$ and $\beta$) for prioritizing these two conditions to generate a cumulative weight which helps selecting the best candidate output. For this purpose, we normalize the TPS (into TPS*) and FIC (into FIC*) with respect to their maximum possible values in the given circuit. Then using $\alpha.$\emph{TPS*} + $\beta.$\emph{FIC*}, we obtain the cumulative weight for the FIC selection process.

\subsection{HSC: Focusing on Higher Skew Gates in FIC}

Our investigation on the hardness of many tested LUT placement strategies revealed that the cells with higher Signal Probability Skew (SPS) at their output are better candidates for obfuscation. The SPS at the output of a gate is defined as $|P_r(0)-P_r(1)|$, with $P_r(1)$ and $P_r(0)$ being the probability of having a 1 or 0 at the output of the gate respectively. The SPS of a gate is a measure of its controllability using primary inputs. The higher the SPS, the lower the controllability of the respective gate. Hence, selecting a high SPS output gate lowers the chances of SAT solver selecting an input that tests the output of that gate. 

\begin{figure*}[t]
\centering
\vspace{-5pt}
\subfloat[]{{\includegraphics[width=0.18\textwidth]{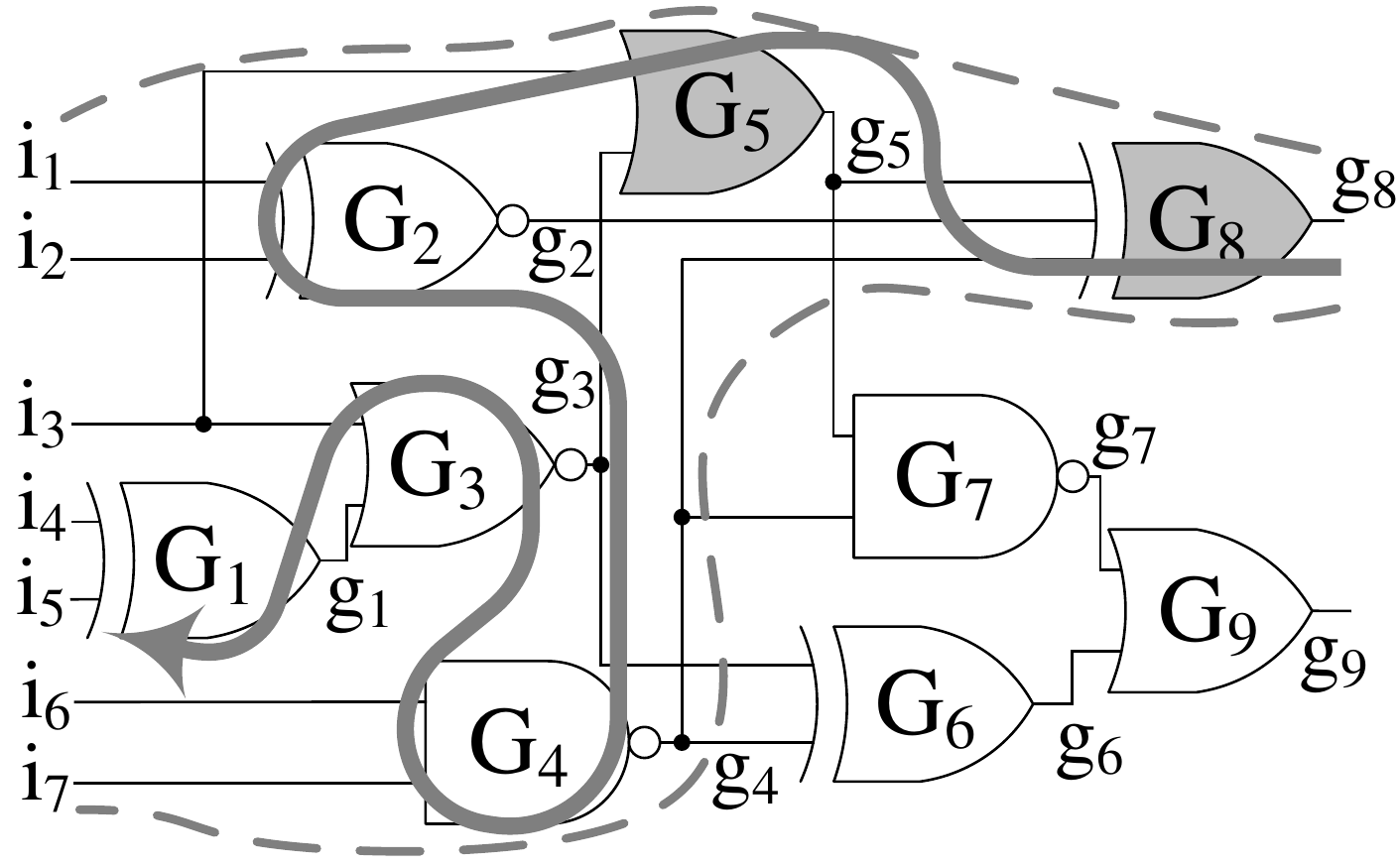} }}%
\subfloat[]{{\includegraphics[width=0.18\textwidth]{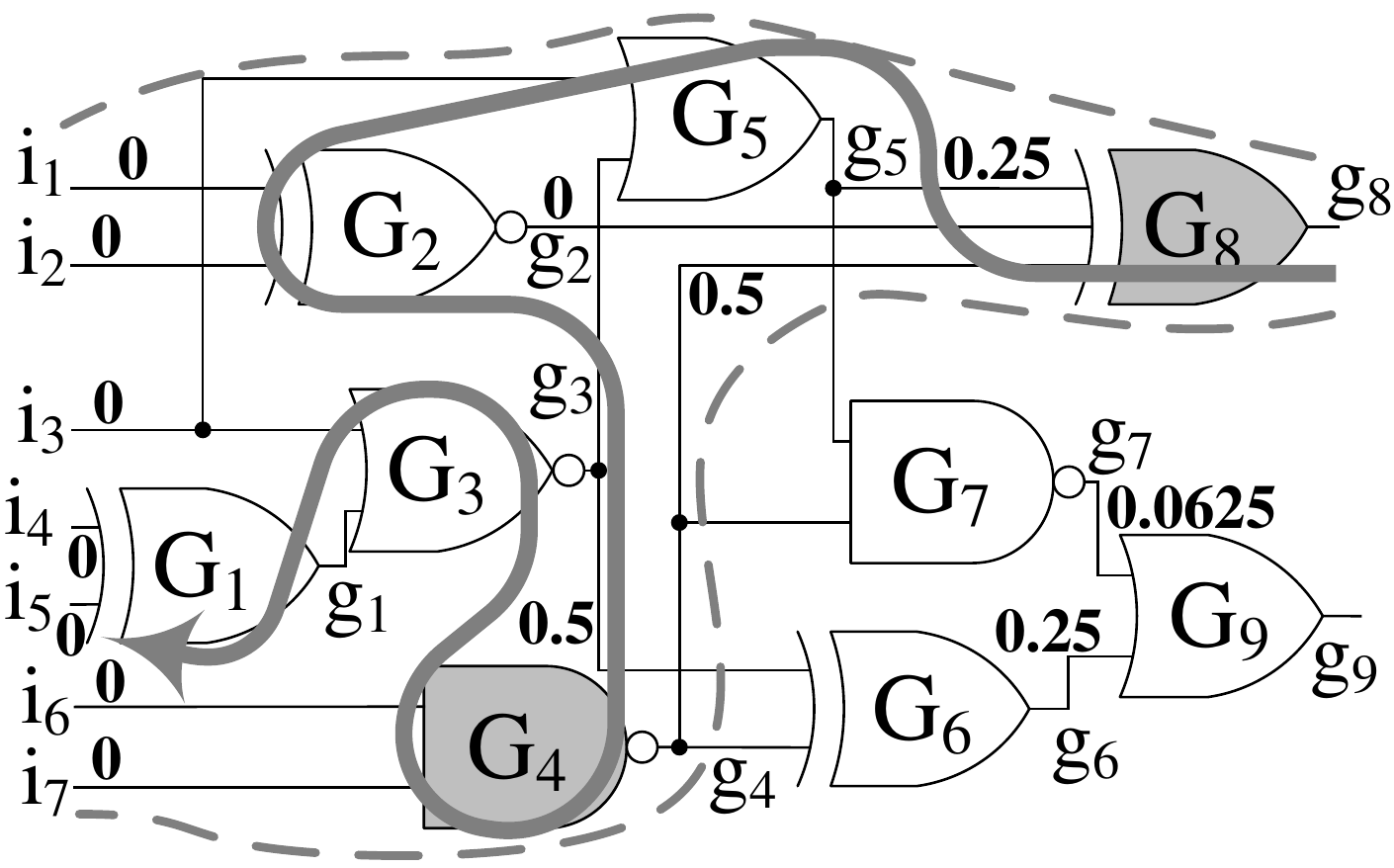} }}%
\subfloat[]{{\includegraphics[width=0.18\textwidth]{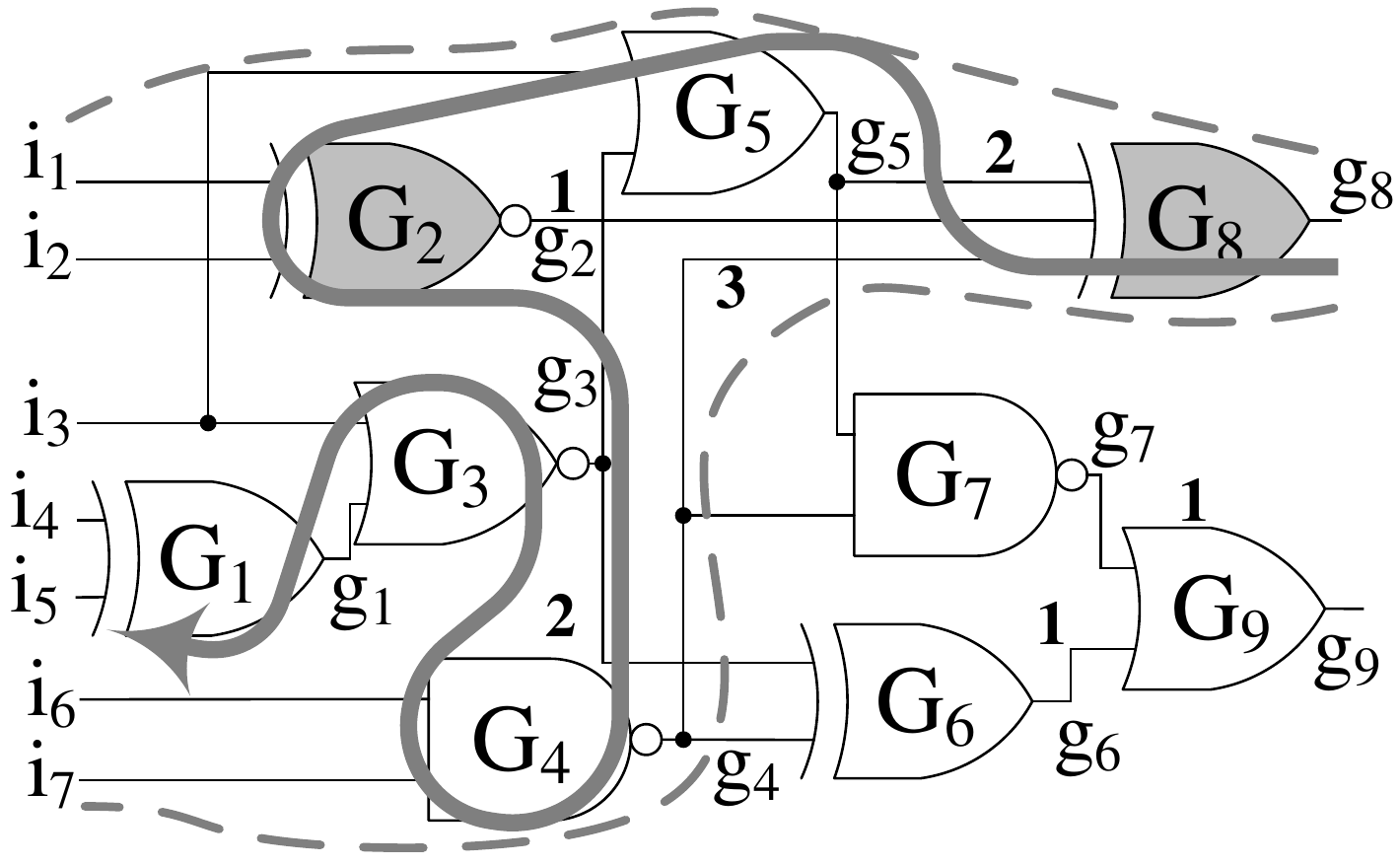} }}%
\subfloat[]{{\includegraphics[width=0.18\textwidth]{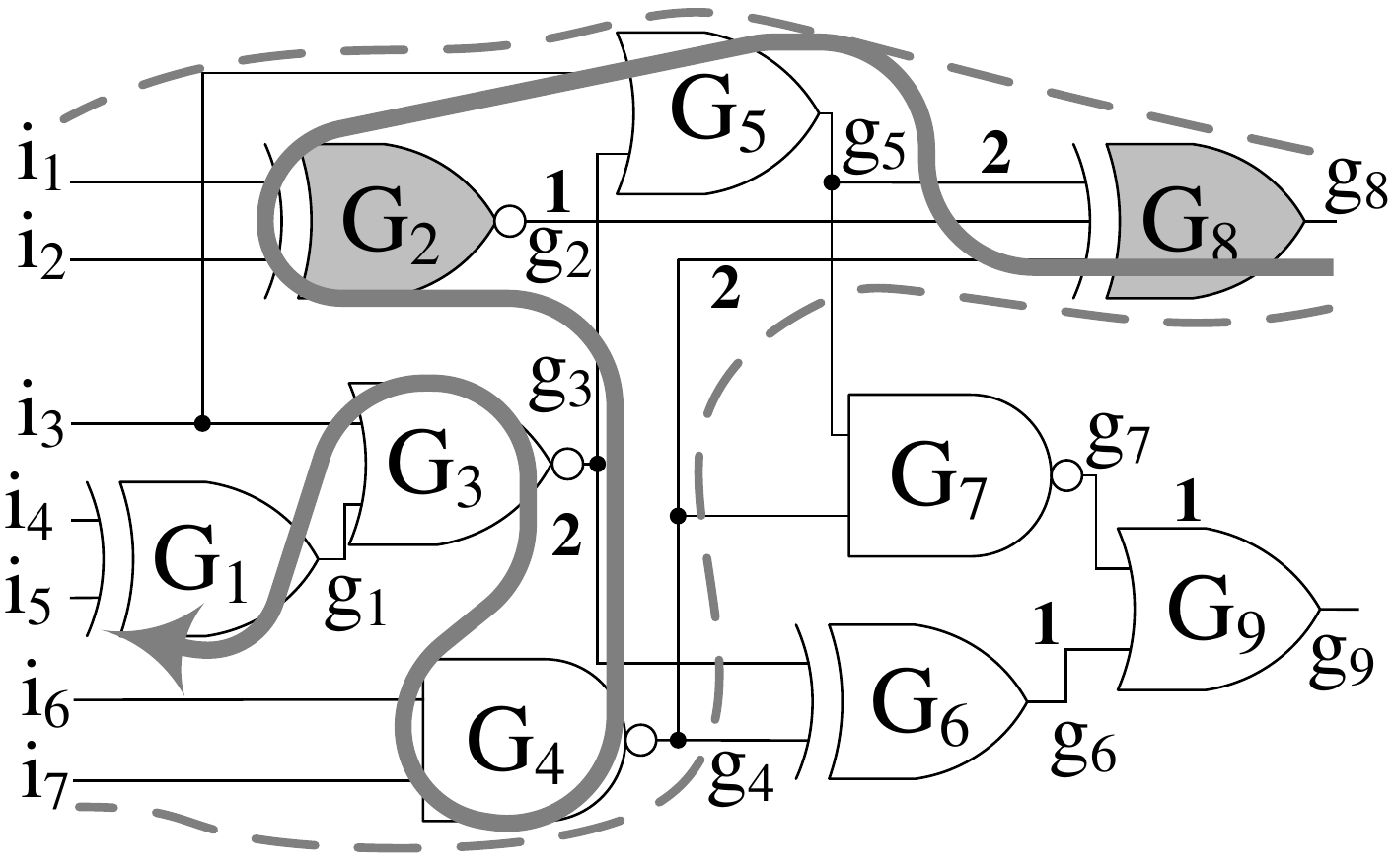} }}%
\subfloat[]{{\includegraphics[width=0.18\textwidth]{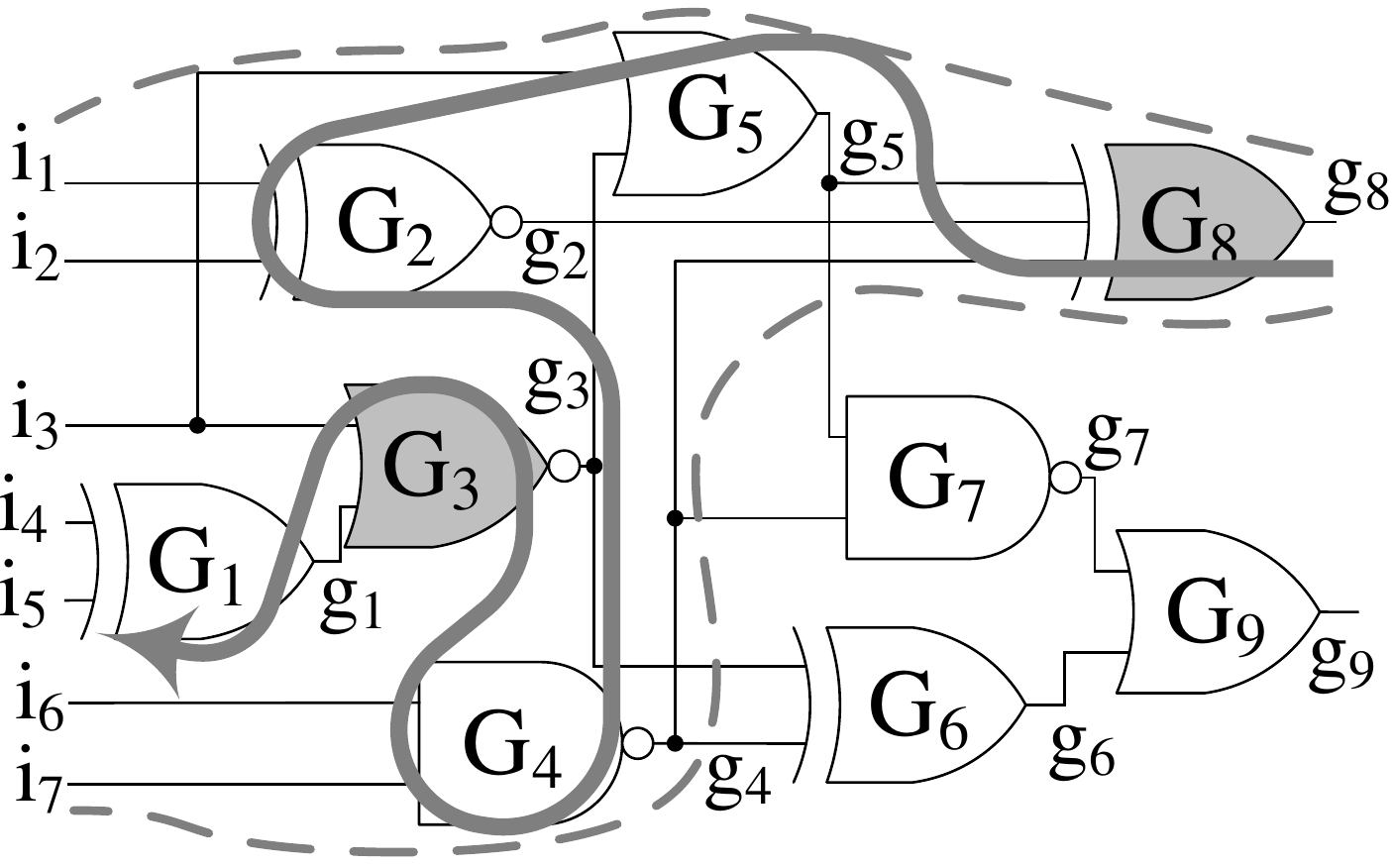} }}%
\vspace{-5pt}
\caption{Gate selection process based on various sub-algorithms: (a) \textbf{FIC}: Focusing on Fan-In Cone of minimum number of primary outputs (b) \textbf{HSC}: Prioritizing higher skew probability gates in FIC (c) \textbf{MFO-HSC}: Prioritizing gates with minimum fan-out \& HSC (d) \textbf{MO-HSC}: Prioritizing gates with minimum impact on outputs \& HSC (e) \textbf{NB2-MO-HSC}: Avoiding back-to-back insertion of LUTs \& MO-HSC. \vspace{-12pt}}
\label{strategies}
\end{figure*}

With this observation, the second step of our proposed algorithm is to modify the FIC to perform the gate selection based on its measure of gate's output (higher) skew probability. In this modified FIC algorithm, which is now referred to as HSC, the gate selection strategy is modified as follows: within the Fan-In cone of selected output(s) based on FIC, the replacement priority is given to gates with higher SPS; In HSC, when a gate is selected for obfuscation, its fan-in gates will be added to the list of gates that could be visited in the next search for gate replacement, and the gates with the highest SPS will be selected among all gates in the list. Similar to FIC algorithm, each gate replacement candidate should pass the timing check, otherwise ignored. HSC replacement flow  is illustrated in Fig. \ref{strategies}(b). In the first invocation of HSC, fan-in cone of gate $G_8$, for satisfying the FIC requirements, is selected and is obfuscated. For the 2$^{nd}$ gate selection, HSC has three candidates $G_2$, $G_5$, and $G_4$. Based on the skew probability of wires, as illustrated in Fig. \ref{strategies}(b),  $G_4$ with SPS of $0.5$ is selected over $G_5$ and $G_2$ with SPS of $0.5$ and \emph{zero} respectively. For the 3rd gate selection, HSC appends the fan-in gates of $G_4$ (Here is primary inputs and will be ignored!) as candidate gates for replacement along with $G_2$ and  $G_5$. Hence, among these 2 gates, $G_5$ is selected for having the higher SPS. 

\vspace{-1pt}
\subsection{MFO-HSC: Focusing on gates with Minimum Fan-Out}

As mentioned previously, lowering the output corruption increases the difficulty of the SAT attack \cite{7495588}\cite{xie2016mitigating}. Although we develop FIC in the first step, the probability of having a fan-in cone with no common gate with other fan-in cones is close to \emph{zero}. Separating the fan-in cones of different outputs could be achieved by replicating common gate, however this will result in a large area overhead. In order to limit the primary output corruption without exploding the area, we introduce another sub-algorithm in which we give obfuscation priority to candidate gate with lowest fan-out. We refer to this gate selection strategy as MFO-HSC.

In MFO-HSC algorithm, a BFS search is first deployed (FIC), visiting all candidate gates at the current breadth, and gate(s) with a minimum number of fan-outs will be selected. Whenever a tie between two or more gates is observed, the gate with the highest SPS is selected. When a gate is obfuscated, its fan-in gates are added to the list of candidate gates that will be visited in the next gate selection. Similar to FIC, each gate replacement candidate should pass the timing check, otherwise ignored. Fig. \ref{strategies}(c) depicts how the MFO-HSC works; Similar to FIC, the fan-in cone of $g_8$ is selected for obfuscation and $G_8$ is obfuscated. Based on BFS, the next candidates are $G_5$, $G_2$, and $G_4$. The gate $G_2$ is selected over $G_5$ and $G_4$ for having fan-out of 1. The fan-in of $G_2$ is then added to the candidate gates for the next visit. In this figure, the fan-in of $G_2$ are primary inputs, and they are ignored, and the the next candidate gate is only $G_5$. 

\vspace{-1pt}
\subsection{MO-HSC: Focusing on Gates with least impact on POs}

Based on our observation in MFO-HSC, there are some gates that have more than one fan-out, but they only affect one output. For instance, as it can be seen Fig. \ref{strategies}(c), the fan-out of $g_4$ is 2. However, it affects only $g_9$. This observation led us to introduce a similar but more efficient sub-algorithm, which is called MO-HSC. In this sub-algorithm rather than looking at the fan-out of the candidate gates, we count the number of outputs that are connected to each candidate gate. MO-HSC requires additional parsing and processing, however it further reduces the output corruption as a result of obfuscation. Similar to MFO-HSC, the tie between two candidate gates (for affecting an equal number of outputs) is broken using SPS of respective gates. Each time a gate is selected for its obfuscation, the fan-in of the gate is added to the list of candidate gates to be considered for the next gate selection. Similar to FIC algorithm, each gate replacement candidate should pass the timing check, otherwise ignored. MO-HSC  is illustrated in Fig. \ref{strategies}(d), where after selecting the $G_8$ based on FIC selection policy, the gate $G_2$ is selected over $G_5$ and $G_4$ for impacting smaller number of outputs.

\subsection{NB2-MO-HSC: Avoiding Back-to-Back insertion of LUTs} \label{demorgan}

\begin{figure}[b]
\centering

\subfloat[]{{\includegraphics[width=46pt]{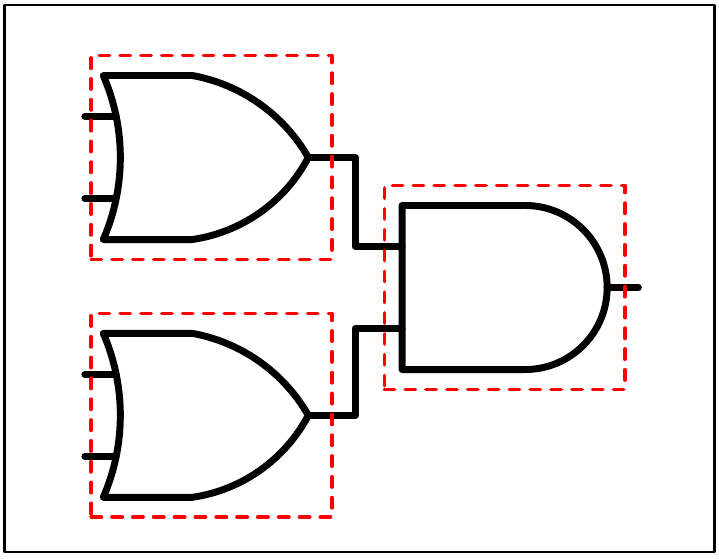} }}%
\subfloat[]{{\includegraphics[width=46pt]{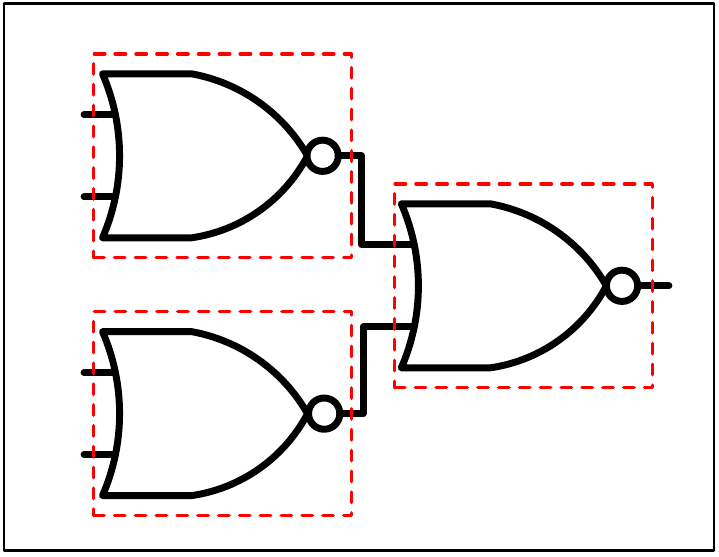} }}%
\subfloat[]{{\includegraphics[width=46pt]{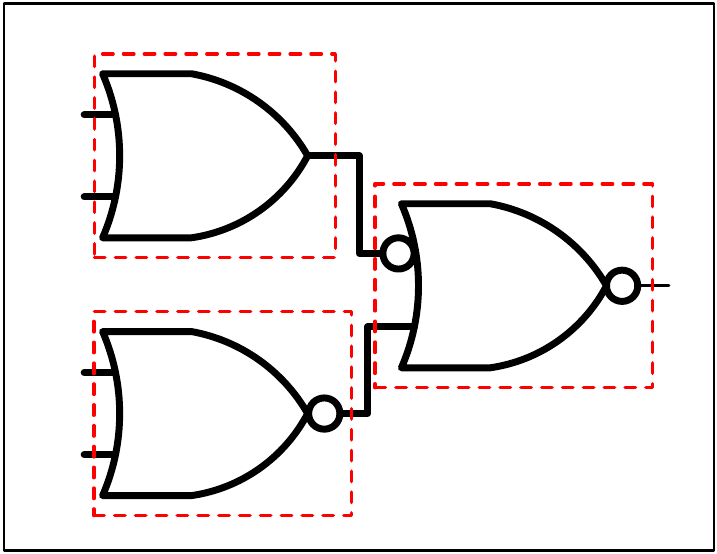} }}%
\subfloat[]{{\includegraphics[width=46pt]{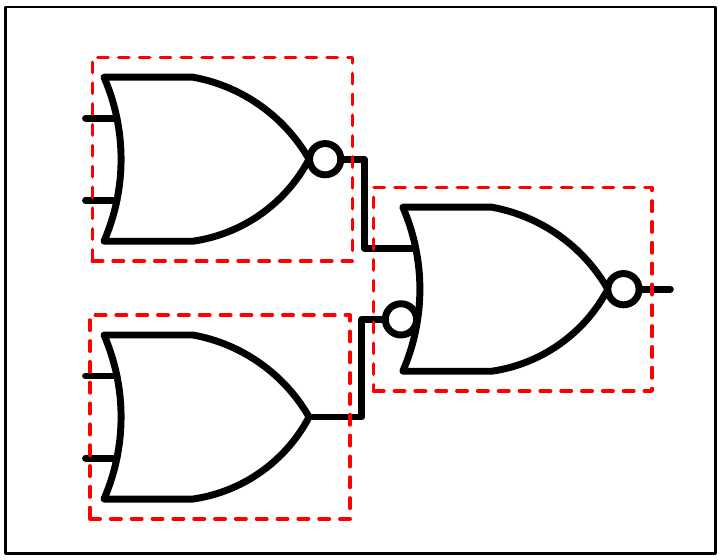} }}%
\vspace{-5pt}
\caption{Gate conversion based on the application of De-Morgan's law (a) OR-AND (b) NOR-NOR (c) Custom1 (d) Custom2.}

\label{fig:Demorgan}
\end{figure}

\begin{figure}[h]
\centering
\includegraphics[width = 210pt]{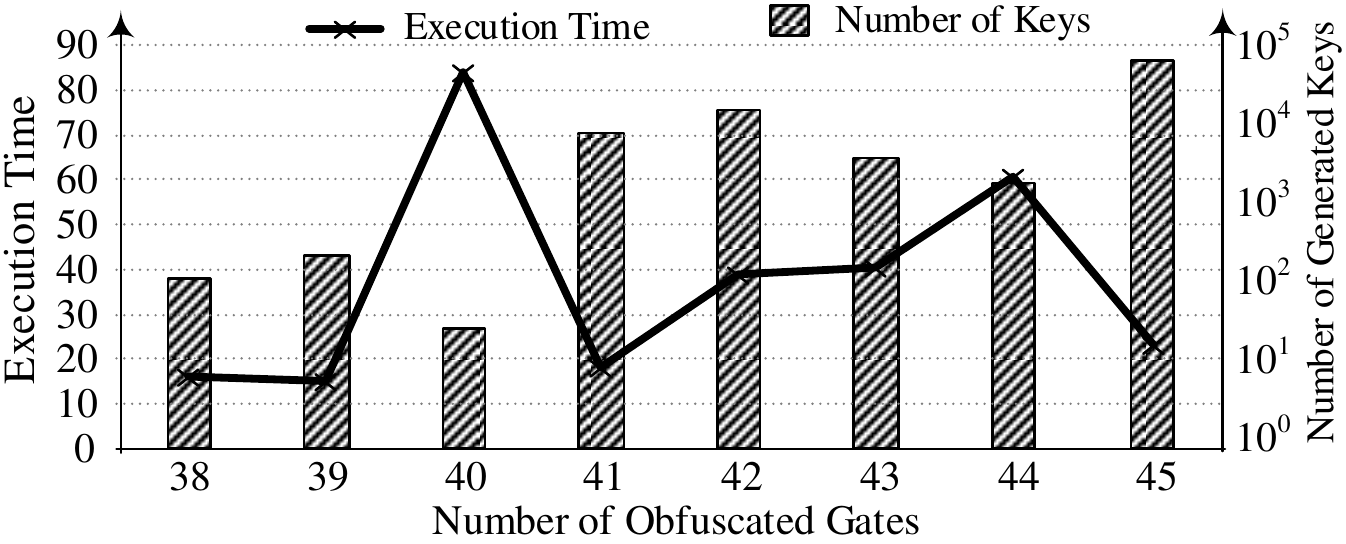}
\vspace{-5pt}
\caption{Increase in the number of valid keys in the result of of back-to-back insertion of LUTs and its impact on SAT attack execution time (C5315).}
\vspace{-8pt}
\label{keyimpact}
\end{figure}

The back-to-back obfuscation of gates with LUTs suffers from the increased number of key-possibilities as a result of the provided freedom in exploiting gate conversion based on De Morgans's Laws. For instance, as it can be seen in Fig. \ref{fig:Demorgan}, the back-to-back obfuscation of the function $(A\lor B)\land (C\lor D)$, using 2-input LUTs, could have 4 different combinations of programmable logic based on De Morgans's Laws. So, the number of correct keys from the intended $1$ increases to $4$. Each additional gate obfuscated in the fan-in of this logic cone, creates another set of possibilities after application of De Morgans's law, leading to exponential increase in the number of valid keys, a phenomenon that we refer to as \emph{correct key explosion}. Depending on the growth rate of the set of valid-keys and the number of keys, obfuscating more gate may even reduce the obfuscation strength. This is illustrated in Fig. \ref{keyimpact} where execution time of a SAT solver, and a number of generated keys per each inserted LUT for the benchmark C5315 of ISCAS-85 is plotted. The LUTs are placed back-to-back, hence, insertion of each LUT increases the number of keys. The plot focuses on the insertion of 38th to the 45th LUT. The insertion of 41st and 42nd LUT, produces a large number of new keys (around $10^4$) based on De Morgan gate conversion possibilities. Hence, the SAT solver execution time doesn't increase. On the other hand, replacement of gate 40 produces far less number of new keys (in range of 10s). Hence the growth of the set of candidate/possible keys exceeds the growth rate of correct keys, significantly increasing the run-time of SAT solver. From this key observation, we need to suppress the growth-rate of correct keys from exploitation of De Morgan's gate conversion laws. So, we introduce another algorithm, NB2-MO-HSC, which implements this restriction by avoiding back-to-back obfuscated, keeping the set of correct keys at a minimum.  In this gate replacement strategy, we first select the candidates in FIC using no back-to-back constraint. Then, the selection among the candidates is made based on candidate gate's connectivity to the minimum number of outputs. If there is a tie among candidates, the SPS of candidate gates determines the selection. As soon as a gate is selected, the NB2-MO-HSC searches the fan-in of the selected gate, skips one logic level (no back to back), and adds the fan-in of all skipped gates to the set of candidate gates for the next gate selection. Similar to FIC, each gate replacement candidate should pass the timing check, otherwise ignored. As illustrated in Fig. \ref{strategies}(e), the application of NB2-MO-HSC results in the selection of $G_8$ and $G_3$ as first two gates to be obfuscated.

The Algorithm \ref{algreplacement} captures the detail implementation of the proposed Lut-Lock obfuscation flow implementing the NB2-MO-HSC policy. As mentioned previously, the overall structures of MFO-HSC and MO-HSC are the same, and since MO-HSC provides slightly more resilient behaviour and also more possible candidates during each iteration, we embed MO-HSC in the proposed LUT-Lock algorithm. 

\algnewcommand\algorithmicforeach{\textbf{for each}}
\algdef{S}[FOR]{ForEach}[1]{\algorithmicforeach\ #1\ \algorithmicdo}

\begin{algorithm}
\caption{LUT-Lock: Implementing NB2-MO-HSC for LUT-based netlist obfuscation}
\begin{algorithmic}[1]
\scriptsize
\State $\alpha$ = $\beta$ = 0.5; \Comment{$\alpha$: TPS coeff,~$\beta$: FIC size coeff};
\State $\gamma$ = 0.1 \Comment{$\gamma$: feasible delay overhead}
\State \emph{max\_delay\_thr} = $\gamma\times$\emph{CriticalPath};
\State MaxSize\_FIC = Max\_TPS = 0; \Comment{Total Positive Slack (TPS)};

\State \emph{Forbidden\_output\_list} = []
\State \emph{outputs\_list} = find\_outputs(\emph{Circuit~C});

\ForEach{(\emph{output} \textbf{in} \emph{outputs\_list})}
    \If{(\emph{output} \textbf{not in} \emph{Forbidden\_output\_list})}
        \State \emph{current\_FIC} = BFS(\emph{output});
        \ForAll {(\emph{paths} \textbf{in} \emph{current\_FIC})}
            \State \emph{Current\_TPS} = TPS\_Calc(\emph{current\_FIC}, paths);
            \State \emph{Current\_Weight} = $\alpha\times$\emph{Current\_TPS} + $\beta\times$sizeof(\emph{current\_FIC})
            \State \emph{Max\_Weight} = $\alpha\times$Max\_TPS + $\beta\times$MaxSize\_FIC
            \If{(Current\_Weight $>$ Max\_Weight)}
                \State \emph{candidate\_output} = \emph{output};
                \State MaxSize\_FIC = sizeof(BFS(\emph{candidate\_output}));
                \State Max\_TPS = \emph{Current\_TPS};
            \EndIf
        \EndFor	
    \EndIf
\EndFor	
\State \emph{candidate\_list} = \emph{Forbidden\_list} = [];
\State \emph{candidate\_list}.append(\emph{candidate\_output});

\While{(\emph{num\_of\_obfuscated} $<$ \emph{target\_no})}
    \If{(\emph{candidate\_list} == $\phi$)}
        \State \emph{Forbidden\_output\_list}.append(\emph{candidate\_output})
        \State \textbf{go to} \emph{line} 5
    \Else
        \State \emph{current\_candidate} = \emph{candidate\_list[0]};
            \If{(delay\_estimate(\emph{current\_candidate}) $<$ \emph{max\_delay\_{thr}})}
                \State replace\_LUT(\emph{current\_candidate});
                \State \emph{current\_candidate\_childlist} = \emph{current\_candidate}.child; 
                \State \emph{Forbidden\_list}.append(\emph{current\_candidate\_childlist});
                \ForEach{(\emph{current\_child} \textbf{in} \emph{current\_candidate\_childlist})}
                    \If{(\emph{current\_child}.child \textbf{not in} \emph{Forbidden\_list})}
                        \State \emph{candidate\_list}.append(\emph{current\_child}.child)
                    \EndIf
                \EndFor
                \State sort\_list(\emph{candidate\_list}, min\_out\_impact);
                \ForAll {(\emph{candidate\_list\_members} \textbf{with} equal min\_out\_impact)}
                \State sort\_list(\emph{candidate\_list}, skew\_probability);
                \EndFor
            \Else
                \State \textbf{remove} \emph{current\_candidate};
            \EndIf
    \EndIf
\EndWhile
\end{algorithmic}
\label{algreplacement}
\end{algorithm}

\section{Experimental Setup} \label{setup}

For benchmarking the proposed LUT-Lock algorithm, we used a farm of desktops equipped with Intel Core-i5 processor and 8GB of RAM. For a fair comparison, and to reduce the impact of the operating system background processes, we dedicated one machine to each SAT solver at a time, and installed Ubuntu Server 16.04.3 LTS operating system in shell mode. We used the largest ISCAS-85~ benchmarks  (C2670, C3540, C5315, C6288, and C7552) to show the effectiveness of the proposed algorithm. We employed the Lingling-based SAT attack described and developed by \cite{7140252}. We measured the SAT solver execution time by increasing the number of obfuscated gates from 1 to 200. A run-time limit of $1.1\times10^4$ seconds was set for the SAT solver.

\section{Results and Discussion} \label{results}

In order to show the effectiveness of each key feature of the proposed algorithm, we compared the execution time of SAT solver on circuits which are obfuscated based on these sub-algorithms. We also compare the effectiveness of the proposed LUT-Lock with that of previous work in STT-LUT \cite{7544331} and Reconfigurable barriers  \cite{baumgarten2010preventing}. 

\begin{figure}[t]
    \centering
    \subfloat[]{{\includegraphics[width=0.94\columnwidth]{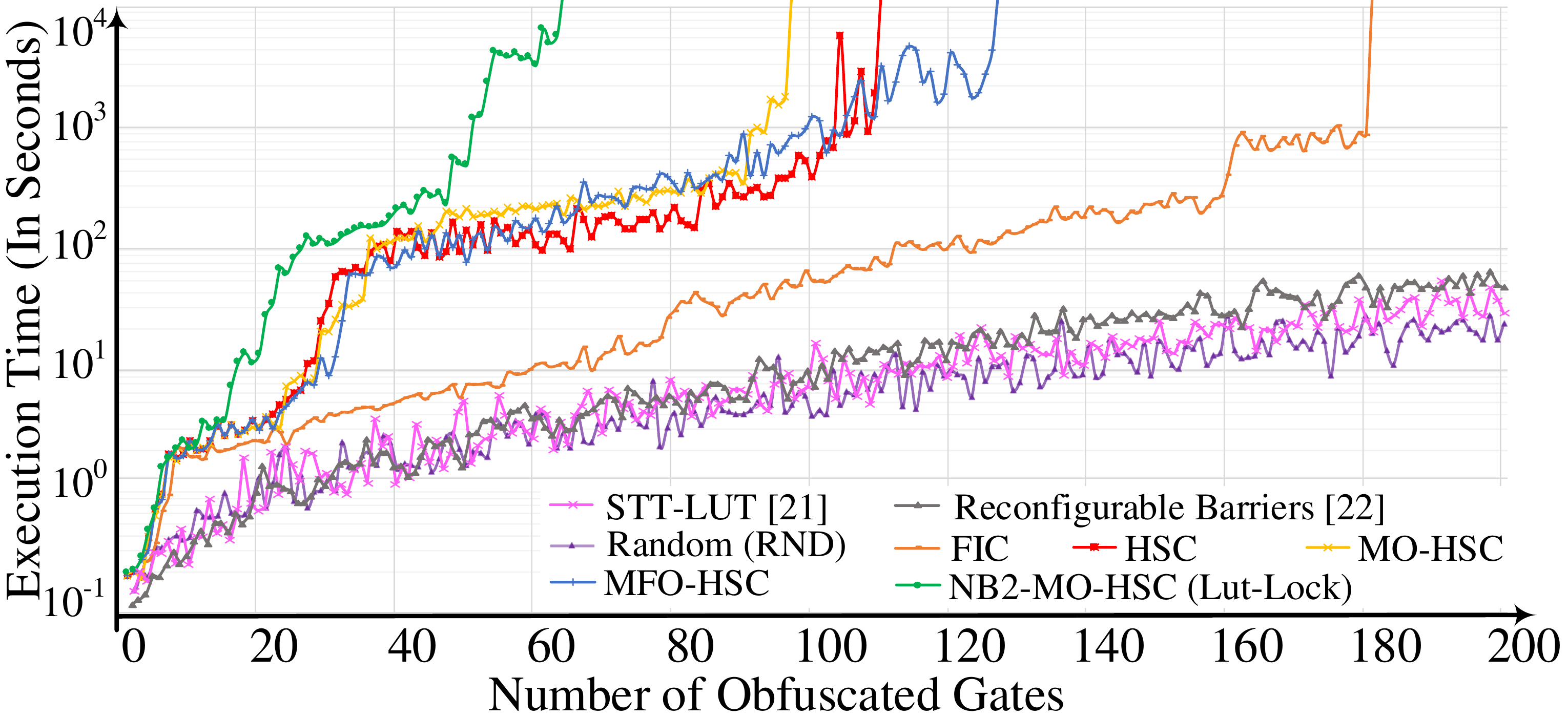}}} \\ 
    \subfloat[]{{\includegraphics[width=0.94\columnwidth]{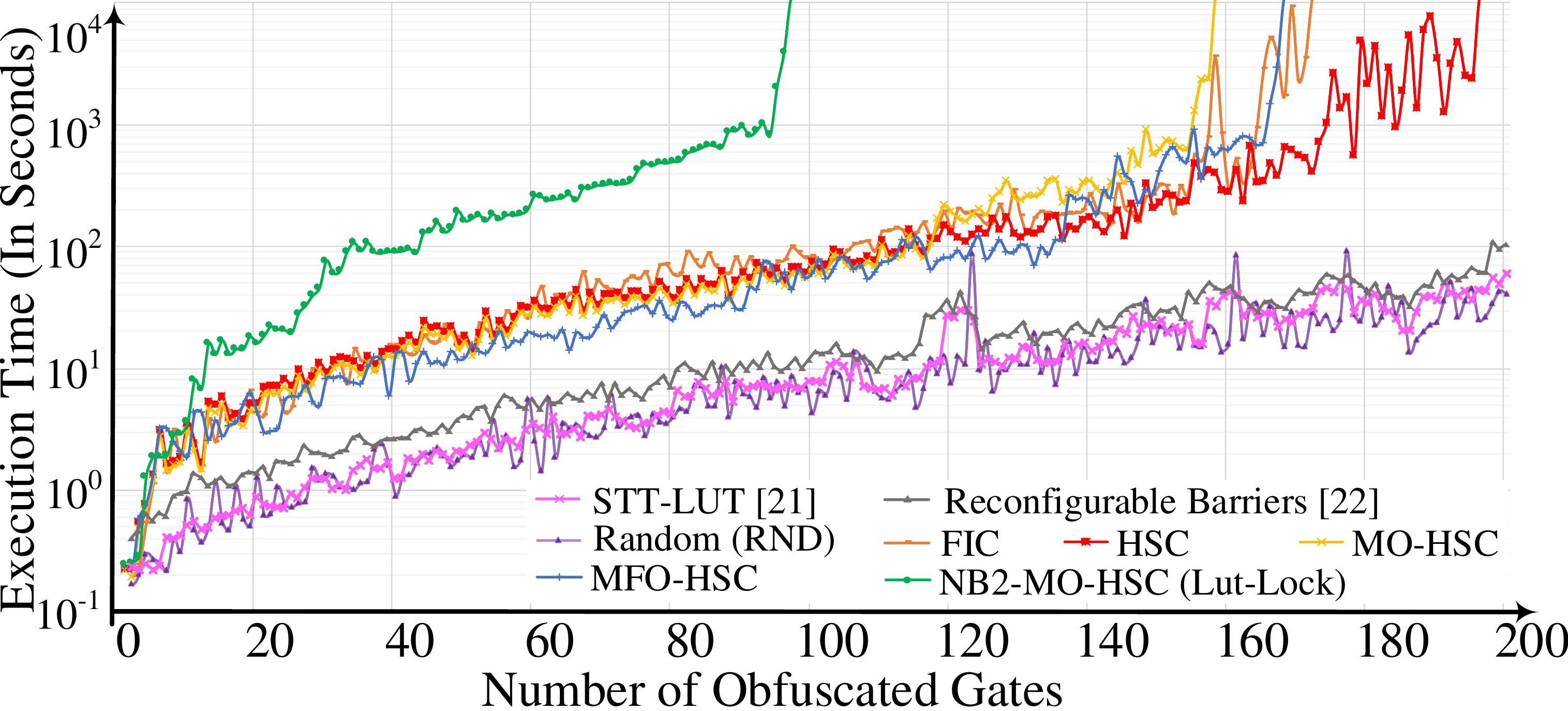}}}
    \vspace{-6pt}
    \caption{Execution time of the SAT solver from \cite{7140252} for finding a valid key when using LUT-Lock (NB2-MO-HSC) compared to its sub-algorithms (RND, FIC, HSC, MO-HSC, and MFO-HSC) and the work in  \cite{7544331} and \cite{baumgarten2010preventing} on ISCAS-85 (a) c5315, (b) c7552 benchmark.}
    \label{LUTFexe}%
\end{figure}

As illustrated in Fig. \ref{LUTFexe} the SAT solver's execution time increases as the replacement algorithm evolves from Random replacement to FIC to HSC to MFO-HSC to MO-HSC to MB2-MO-HSC, illustrating the orthogonal improvement of added features in providing resiliency against SAT attacks. The LUT-Lock algorithm, implementing the NB2-MO-HSC replacement policy, combines all key features and provides a close to exponential increase in the execution time of SAT attack with respect to the number of obfuscated gates. 

As illustrated in Fig. \ref{LUTFexe}, the execution time of the SAT solver, although increases steadily, faces small variation. The variation in the execution time is the result of (1) random nature of SAT solver in selecting DIPs from run-to-run, and (2) rate of growth in the size of valid keys (as a result of gate conversion using the application of De Morgan's laws, as explained in section \ref{demorgan}), compared to the rate of growth in the number of possible keys. A poor selection of candidates for obfuscation results in a faster growth in the number of valid keys, reducing the overall effectiveness of obfuscated netlist against the SAT attack. As illustrated, the LUT-Lock has the least variation, as it eliminates the explosion of the set of valid keys by preventing back-to-back gate obfuscation. 

Table \ref{averageexe} captures the fitted function of execution time for different sub-algorithms and LUT-Lock, where x denote the number of obfuscated gates. As illustrated in this Table, the LUT-Lock (NB2-MO-HSC) poses an exceptionally more challenging SAT problem compare to other obfuscation scheme. Table \ref{averageexeperc} compare the execution time of SAT attack, across selected number of ISCAS 85 benchmarks, once obfuscated by random LUT insertion and once using LUT-Lock. As illustrated, execution time of SAT attack, grows slowly as number of obfuscated cells increases in random insertion policy, but it grows exponentially when LUT-Lock policy is adopted.

Figure \ref{LUTFexe} visualizes the growth in the execution time of SAT attack, for two of ISCAS-85 benchmarks obfuscated using various LUT replacement policies. Other benchmarks have similar behaviour and are omitted for lack of space. In addition to replacement policies discussed in this paper, the SAT resiliency of replacement policies in prior work, namely  STT-LUT \cite{7544331} and Reconfigurable barriers  \cite{baumgarten2010preventing} are captured in this figure. From this figure, the SAT resiliency of prior work is close to that of random replacement, showing slow growth in SAT attack execution time with respect to the number of inserted gates, where the Lot-Lock replacement policy clearly shows a much faster exponential increase in difficulty. As illustrated, in both benchmarks, with only 20 replaced LUTs, the LUT-Lock obfuscated netlist is as resilient as the netlist produced by \cite{7544331} and \cite{baumgarten2010preventing} replacement policy when using $10X$ (200 gates) the number of gates. And by increasing the number of gates, the SAT resiliency of LUT-Lock insertion policy still grows exponentially.     

\begin{table}[t]
\centering
\caption{Average Execution Time of SAT attack across studied benchmarks obtained based on curve fitting, where $x$ is the number of obfuscated gates.}
\vspace{-8pt}
\includegraphics[width = \columnwidth]{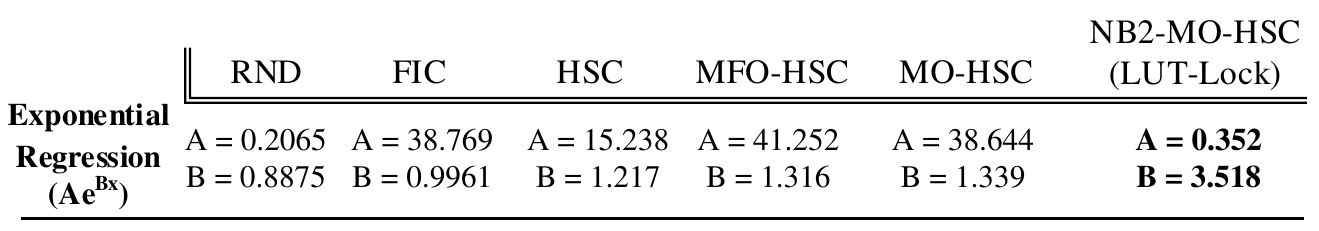}
\vspace{-14pt}
\label{averageexe}
\end{table}

\begin{table}[t]
\centering
\caption{Average Execution Time of SAT Solver across studied benchmarks as a function of the percentage of obfuscated gates.}
\vspace{-6pt}
\includegraphics[width = \columnwidth]{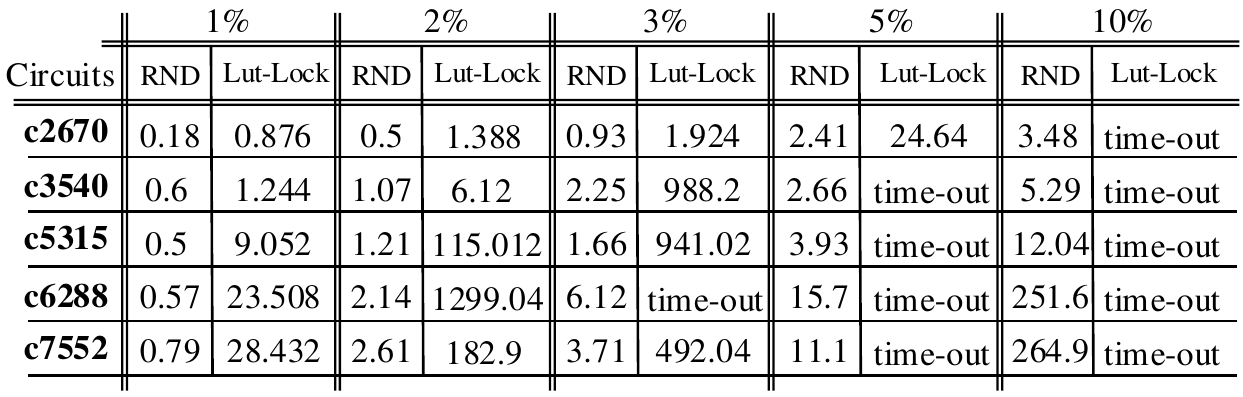}
\vspace{-12pt}
\label{averageexeperc}
\end{table}

\vspace{-1pt}
\section{Conclusions} \label{conclusion}
\vspace{-1pt}

We proposed the LUT-Lock, a novel LUT-based obfuscation algorithm, for building SAT resilient obfuscation netlists, applicable to FPGA and ASIC designs. Our simulation results illustrated that focusing the obfuscation to impact the smallest number of primary output pins increases the obfuscation difficulty. This was achieved by means of selecting the fan-in of a minimum number of primary outputs for obfuscation and selecting gates that are connected to the smallest number of output pins. In addition, we illustrated that gates with lower controllability with respect to the primary inputs, as measured by signal probability skew at their output pin, are better candidates for obfuscation. Furthermore, we illustrated that avoiding back-to-back LUT replacement considerably reduces the number of valid key possibilities, increasing the resiliency of the proposed algorithm against SAT attacks. Compared to previous work, the LUT-Lock (NB2-MO-HSC) algorithm provides exponentially better protection against SAT attacks.

\end{document}